\documentclass[]{aa}
\def\apj{ApJ}

\def\aap{A\&A}

\def\ueber#1#2{{\setbox0=\hbox{$#1$}%
  \setbox1=\hbox to\wd0{\hss$ #2$\hss}%
  \offinterlineskip
  \vbox{\box1\box0}}{}}

\def\lesssim{\,\lower 1mm \hbox{\ueber{\sim}{<}}\,}
\def\grsim{\,\lower 1mm \hbox{\ueber{\sim}{>}}\,}

\makeatletter

\let\@internalcite\cite
\def\cite{\@ifstar{\citeyear}{\citefull}}
\def\citefull{\def\astroncite##1##2{##1 ##2}\@internalcite}
\def\citeyear{\def\astroncite##1##2{##2}\@internalcite}

\def\citeau{\def\astroncite##1##2{##1}\@internalcite}
\def\citen{\def\astroncite##1##2{##1 (##2)}\@internalcite}
\def\possesivcite{\def\astroncite##1##2{##1's (##2)}\@internalcite}

\def\@citex[#1]#2{\if@filesw\immediate\write\@auxout{\string\citation{#2}}\fi
  \def\@citea{}\@cite{\@for\@citeb:=#2\do
    {\@citea\def\@citea{; }\@ifundefined
       {b@\@citeb}{{\bf ?}\@warning
       {Citation `\@citeb' on page \thepage \space undefined}}%
{\csname b@\@citeb\endcsname}}}{#1}}
\def\@cite#1#2{#1\if@tempswa , #2\fi}
\def\@biblabel#1{}
\makeatother

\title{The gravitational collapse of ONe electron-degenerate cores 
and white dwarfs: the role of $^{24}$Mg and $^{12}$C revisited}

\author{Jordi Guti\'errez\inst{1},
        Ramon Canal\inst{2,3,4} \and
        Enrique Garc\'{\i}a--Berro\inst{1,3}}
\titlerunning{Gravitational collapse of ONe cores}
\authorrunning{J. Guti\'errez, et al.}

\institute{$^1$Departament de F\'\i sica Aplicada, Escola  Polit\'ecnica 
               Superior de Castelldefels, Universitat   Polit\`ecnica de 
               Catalunya,   Avda.  del  Canal  Ol\'\i  mpic  s/n,  08860 
               Castelldefels, Spain\\
           $^2$Departament   d'Astronomia  i   Meteorologia, Universitat 
               de Barcelona,   Facultat  de   F\'\i sica,  Mart\'{\i}  i 
	       Franqu\`es 1, 08028 Barcelona, Spain\\
           $^3$Institute for Space Studies of Catalonia, c/Gran Capit\`a
               2--4, Edif.  Nexus 104, 08034 Barcelona, Spain\\
           $^4$Special Research Center in Astrophysics, Particle Physics
               and  Cosmology, University  of  Barcelona,  Mart\'{\i}  i 
               Franqu\`es 1, 08028 Barcelona, Spain\\
           }

\offprints{E. Garc\'\i a--Berro}
\date{\today}

\abstract{The  final stages of  the evolution  of electron--degenerate
ONe  cores,   resulting  from  carbon  burning   in  ``heavy  weight''
intermediate--mass stars ($8 \, M_{\sun}\la M \la 11 \, M_{\sun}$) and
growing  in  mass, either  from  carbon burning  in  a  shell or  from
accretion  of matter in  a close  binary system,  are examined  in the
light of  their detailed chemical composition. In  particular, we have
modelled  the evolution  taking  into account  the  abundances of  the
following  minor  nuclear  species,  which result  from  the  previous
evolutionary    history:   $^{12}$C,    $^{23}$Na,    $^{24}$Mg,   and
$^{25}$Mg. Both  $^{23}$Na and $^{25}$Mg give rise  to Urca processes,
which  are  found to  be  unimportant for  the  final  outcome of  the
evolution. $^{24}$Mg was formerly  considered a major component of ONe
cores   (hence   called  ONeMg   cores),   but  updated   evolutionary
calculations in  this mass range have severely  reduced its abundance.
Nevertheless,  we have  parameterized it  and we  have found  that the
minimum  amount  of  $^{24}$Mg  required  to produce  NeO  burning  at
moderate densities  is $\sim  23\%$, a value  exceedingly high  in the
light of recent evolutionary models.  Finally, we have determined that
models   with   relatively   small   abundances  of   unburnt   carbon
($X(^{12}$C)$\sim 0.015$)  could be a  channel to explosion at  low to
moderate  density ($\sim 1\times  10^9$~g~cm$^{-3}$). This  is clearly
below the  current estimate  for the explosion/collapse  threshold and
would have interesting consequences.
 }

\begin{document}

\maketitle

\section{Introduction}

The final  evolution of electron-degenerate  cores made of  oxygen and
neon (hereafter  ONe) has  received little attention  in the  last few
years despite  its potential importance for supernova  theory.  One of
the reasons  for this lack of  accurate models is that  there are very
few realistic  evolutionary calculations  leading to the  formation of
such cores.   This, in turn,  is due to  the fact that for  stars with
masses ranging  from 8 to  $11\, M_{\sun}$, carbon-burning  is ignited
off-center and proceeds through a  series of mild flashes in partially
degenerate conditions,  a situation especially  difficult to simulate,
which requires very short time-steps  and careful zoning, leading to a
heavy computational demand.

Up to now, no  detailed multidimensional calculations of the evolution
of   high--end  intermediate   mass   stars  in   this  phase   exist.
Nevertheless,   there   are   several   possible  effects,   such   as
rotation--induced mixing and  overshooting, and, more importantly, the
possibilty  of  off--center ignitions  (Garc\'\i  a--Senz \&  Woosley,
1995; Woosley,  Wunsch \& Kuhlen,  2004; Wunsch \& Woosley  2004) that
should be modeled  and discussed using multi-dimensional hydrodynamics
--- see,  however, H\"oflich \&  Stein (2002).   Some multidimensional
stellar  models already  exist,  but they  are  in an  early stage  of
development.  Spherically--symmetric models include a degree of detail
that current  two-- and  three--dimensional simulations are  still far
from reaching. A discussion of this kind of effect is beyond the scope
of  this   paper.  However,  although   multidimensional  evolutionary
calculations of stars in this mass  range do not exist yet, Hirschi et
al. (2004)  have studied the  evolution of rotating stars  with masses
$M\geq 12\,M_{\odot}$  --- slightly larger than  those considered here
--- and  have   found  that  most  of  the   differences  between  the
presupernova  structures  obtained  from  rotating  and  non--rotating
stellar models have their origin in the effects of rotation during the
core   hydrogen  and  helium   burning  phases.    Additionally,  they
explicitly  mention  that  the  advanced stellar  evolutionary  stages
appear too short in time to allow the rotational instabilities to have
a significant impact during the late stages.  Hence, given the lack of
accurate multidimensional calculations for this phase of the evolution
of  heavy--weight intermediate  mass  stars, one  should conclude,  in
principle,  that spherically--symmetric evolutionary  calculations can
still provide a reasonable approximation  to the real behavior --- see
the recent review of Maeder \& Meynet (2000) for a thorough discussion
of this topic.

These  objects were  long ago  proposed  as clear  candidates for  the
so--called accretion--induced  collapse --- a  process first described
by Canal \&  Schatzman (1976).  Currently, it is  widely accepted that
ONe  electron-degenerate objects do  collapse to  form a  neutron star
when the central density is driven beyond a certain threshold --- see,
for  instance,   Guti\'errez  et  al.   (1966,  1997),   or  Canal  \&
Guti\'errez  (1997), and  references therein.   In particular,  in the
pioneering work of  Miyaji et al.  (1980) it was  shown that all stars
in  the mass interval  $8 \,  M_{\sun}\la M\la  12 \,  M_{\sun}$ would
develop electron--degenerate ONeMg  cores during carbon shell burning.
These cores would undergo a phase of electron capture on $^{24}$Mg and
$^{24}$Na  first, and  later  on $^{20}$Ne  and  $^{20}$F, to  finally
ignite Ne  and O  explosively at central  densities higher  than $\sim
2\times10^{10}$~g~cm$^{-3}$.   At these  very high  central densities,
fast electron captures occuring on the nuclear statistical equilibrium
(NSE) material  would rapidly drive  the Chandrasekhar mass  below the
actual mass  of the  degenerate core and,  consequently, gravitational
collapse  would ensue.   Later,  Woosley, Weaver  \&  Taam (1980)  and
Nomoto (1984) reduced the above mass  range to $8 \, M_{\sun} \la M\la
10 \, M_{\sun}$.  Stars more  massive than $\sim 10 \, M_{\sun}$ would
ignite Ne in a series  of non-explosive flashes and ultimately proceed
through the O--burning and Si--burning stages in a way similar to more
massive  stars.  Subsequently,  Timmes \&  Woosley (1992,  1994) found
that the density threshold for an ONe core to collapse to neutron star
dimensions  was  around   $\sim  4\times  10^9$~g~cm$^{-3}$.   On  the
contrary, if NSE was reached at densities below this critical density,
the  degenerate  object   would  be  completely  disrupted.   Finally,
Guti\'errez et  al.  (1996) analyzed the role  of Coulomb corrections,
both in the  equation of state and in  the electron--capture threshold
energies,  and  found  that  explosive  NeO ignition  takes  place  at
densities  high enough  to  ensure gravitational  collapse to  nuclear
matter densities.  The previously  described scenario should apply not
only  to the  cores  of Super--Asymptotic  Giant  Branch (SAGB)  stars
(Garc\'{\i}a--Berro  \& Iben  1994)  but also  to  ONeMg white  dwarfs
accreting material from  a companion in a close  binary system (Miyaji
et al.  1980; Nomoto 1984, 1987).

Most of  the studies done  so far for  this mass range  had considered
$^{24}$Mg as a  major constituent of the core.   However, the detailed
studies    of   Ritossa,    Garc\'{\i}a--Berro    \&   Iben    (1996),
Garc\'{\i}a--Berro,   Ritossa  \&  Iben   (1997),  Iben,   Ritossa  \&
Garc\'{\i}a--Berro  (1997)  and  Ritossa, Iben  \&  Garc\'{\i}a--Berro
(1999) have clearly shown that the  mass fraction of $^{24}$Mg is much
smaller than  previously thought. Following the trends  shown by these
modern evolutionary calculations for the  stars in this mass range, in
this paper we  investigate the effects of the  chemical composition of
the  degenerate core  on the  possible outcome  of these  objects.  In
particular, we analyze  the impact of the abundance  of $^{24}$Mg, the
presence  of $^{23}$Na  and  $^{25}$Mg, and  the  presence of  unburnt
$^{12}$C,  which appears to  be a  specific characteristic  of current
evolutionary models. The paper is  organized as follows.  In section 2
we describe  the most  relevant input physics  and the  initial model.
Section  3  is devoted  to  the  effects  of an  increasing  $^{24}$Mg
abundance, while in \S 4 we  analyze the role played by the previously
unburnt $^{12}$C.   The role of the minor  chemical species ($^{23}$Na
and $^{25}$Mg) is analyzed in \S 5. In \S 6 we discuss our results and
we draw our conclusions.

\section{Input physics and initial model}

For the  sake of  conciseness, and given  that the numerical  code and
most of the physical inputs  employed in the calculations discussed in
this  paper have  already been  described in  previous papers  --- see
Guti\'errez  et al.  (1996) and  references  therein ---  we will  not
present  them  here  in  detail.  Instead, in  this  section  we  only
summarize the  most relevant  physical ingredients which  are relevant
for the discussion of the evolutionary sequences analyzed below. These
are  the   electron  capture   rates,  Coulomb  corrections   and  the
prescription adopted for convective transport.

Electron capture rates  have been taken from Oda  et al. (1994), which
provides    the   most    up-to-date,   and    physically   consistent
determinations.  This point  seems  already fixed,  as  the rates  for
sd-shell  nuclei have  remained  almost unchanged  since  the work  of
Takahara et  al. (1989). However, the  data of Oda et  al.  (1994) are
much  richer  for  $^{24}$Mg,  $^{20}$Ne  and  other  sd-shell  nuclei
relevant for this work, such as $^{23}$Na and $^{25}$Mg.

Coulomb  corrections  to  the  electron capture  threshold  have  been
included as in Couch and Loumos  (1974). The net effect on the Coulomb
interaction energy is to lower it, hence reducing the Fermi energy and
increasing  the  capture   density  threshold.  The  exact  expression
to compute the energy shift is

\begin{eqnarray}
\Delta E_{\rm thr} = \mu(Z+1)-\mu(Z)
\end{eqnarray}

\noindent where $\mu(Z)$  is the chemical potential of  the nucleus of
charge $Z$, given by

\begin{eqnarray}
\mu(Z) = -&k_{\rm B}&T\Big(\frac{Z}{{\bar Z}}\Big) 
\Bigg\{ \Gamma \Big[ 0.9 +
c_1\Big(\frac{Z}{\bar Z}\Big)^{1/3} + c_2
\Big(\frac{Z}{\bar Z}\Big)^{2/3}\Big] + \nonumber \\
\Big[&d_0& + d_1 \Big(\frac{Z}{\bar Z}\Big)^{1/3} + ... \Big]
+ O(1/\Gamma) \Bigg\}
\end{eqnarray}

\noindent  where  $\Gamma  =  {\bar Z}^{5/3}e^2/a_{\rm  e}  k_{\rm  B}
T=2.275\times  10^5 {\bar  Z}^{5/3} (\rho  Y_{\rm e})^{1/3}/T$  is the
Coulomb  coupling  parameter,  $a_{\rm  e}$  is  the  inter-electronic
distance, $\bar Z$ is the average charge, and the four constants adopt
the  following  values:  $c_1=0.2843$, $c_2=-0.054$,  $d_0=-9/16$  and
$d_1= 0.460$.  The  rest of the symbols have  their usual meaning. All
the electron captures, as well as the Urca processes, are considerably
affected by  the change  in the  threshold.  In the  case of  the Urca
processes,  the  circumstances  are especially interesting  since  the
electron capture threshold will  be increased, while the threshold for
the $\beta$  decay will decrease.   At the densities  and temperatures
relevant for this study, the change in the threshold energy amounts to
roughly $\sim  300$~keV, or  to an electron-capture  density threshold
increase  of  $\sim  5\times  10^8$~g~cm$^{-3}$,  thus  favouring  the
collapse.

Another  very important  issue is  what convective  criterion  must be
adopted. In the paper of  Miyaji et al. (1980), convection was assumed
to set in according to  the Schwarzschild criterion: a convective core
started  to  develop  at  the   onset  of  the  electron  captures  on
$^{24}$Mg.   Convective  heat  transport   contributed  to   keep  the
temperature  below  that  of  explosive  Ne  ignition  ($\sim  2\times
10^{9}$~K) along  core contraction until  a central density  $\rho \ga
2\times  10^{10}$~g~cm$^{-3}$  was reached.   That  was questioned  by
Mochkovitch (1984), who argued that the steep gradient of the electron
mole number, $Y_{\rm e}$, produced  by the same electron captures that
create the  superadiabatic temperature gradient,  should stabilize the
fluid  against convective  motion  and a  semiconvective region  would
form.   Miyaji \&  Nomoto  (1987) tested  this  situation by  assuming
semiconvective mixing to be  negligible. They found that explosive NeO
burning   was   ignited  at   a   central   density  $\sim   9.5\times
10^{9}$~g~cm$^{-3}$: heating  by the electron captures  and cooling by
thermal neutrinos  was then  purely local.  Thus,  convective criteria
are essential  in ascertaining the  final outcome of the  evolution of
compact  ONe objects.   In  this  paper we  adopt  a rather  pragmatic
approach.   That is,  we  first consider  the classical  Schwarzschild
criterion $\nabla \le \nabla_{\rm ad}$.  However, since this stability
criterion assumes  complete chemical homogeneity  and this is  not the
case when  electron-captures start, we also use  the Ledoux criterion:
$\nabla \le \nabla_{\rm L}$, where:

\begin{equation}
\nabla_{\rm L} \equiv \nabla_{\rm ad} -
\left[ \left(\frac{\partial \ln P}{\partial \ln Y_{\rm e}}\right)_T\Bigg/
\left(\frac{\partial \ln P}{\partial \ln T}\right)_{Y_{\rm e}} \right]
\nabla_{Y_{\rm e}}
\end{equation}

To account  for the convective mixing,  we have employed  a simple but
robust  numerical procedure.  When  convection sets  in, the  chemical
composition is homogeneized in the region given by

\begin{eqnarray}
r_{\rm mix} = \min (\Lambda_{\rm mix,k}, v_{\rm conv,k}\times \Delta t)
\end{eqnarray}

\noindent  where  $\Lambda_{\rm mix,k}$  and  $v_{\rm  conv, k}$  are,
respectively, the mixing length scale and the convective velocity of a
given mass shell and $\Delta t$  is the time-step. So, we are assuming
a high  convective efficiency in  both chemical transport  and entropy
transport.

Our  initial model,  used  in all  the  calculations, is  a $1.375  \,
M_{\sun}$ electron-degenerate object, whose main chemical constituents
are $^{16}$O and $^{20}$Ne,  with the distribution obtained by Ritossa
et al. (1996).  Its central density and temperature are, respectively,
$\rho_{\rm  c}  \simeq 1.0  \times  10^9$~g~cm$^{-3}$  and $T_{\rm  c}
\simeq 2.0\times 10^8$~K.  The core  is discretized in up to 1000 mass
shells by an adaptive algorithm, which ensures good resolution both in
chemical composition and in entropy throughout the structure.  The ONe
object is slowly  contracting due to spherical accretion  at a rate of
$\sim 10^{-6}\, M_{\sun}$~yr~$^{-1}$.  Its mass is consistent with the
mass  of ONe white  dwarfs, and  is close  to the  Chandrasekhar mass,
making this  object prone to experience sudden  instability leading to
either disruption or collapse.

\section{The effects of the abundance of $^{24}$Mg}

Since the pioneering work of  Miyaji et al.  (1980), the astrophysical
nuclear  reaction  rates   have  undergone  significant  changes.   In
particular, the amount  of $^{24}$Mg present in typical  ONe cores has
been  reduced  by  almost  an  order  of  magnitude  (Ritossa  et  al.
1996). Currently, it  seems that the most important  rates are already
reasonably  well known but,  nonetheless, it  would be  interesting to
ascertain  the minimum  abundance of  $^{24}$Mg required  to  cause an
early   explosion  of  the   degenerate  core.   For  this,   we  have
parameterized the  abundance of  $^{24}$Mg while keeping  constant the
ratio O/Ne.  While this  is not truly  consistent, it serves  well the
point relevant  here: how  much $^{24}$Mg is  the minimum to  engage a
explosion  at about $4\times  10^9$~g~cm$^{-3}$, which  would possibly
lead to  the complete disruption  of the degenerate object.   With the
currently accepted  abundances of $^{24}$Mg $(\sim 3\%$  by mass), the
maximum  temperature attained  during  the electron  captures is  only
$3.8\times  10^8$~K (Guti\'errez  et  al. 1996),  very  far below  the
minimum needed to  initiate the explosive oxygen burning  and to drive
the  matter to  NSE.  The  goal of  this section  is to  determine the
minimum  abundance   of  $^{24}$Mg  required  to   produce  a  central
temperature of about $1.5 \times 10^9$~K.

In  Fig.~1 several evolutionary  sequences with  increasing abundances
(3\%, 10\%, 15\% and 25\%) of $^{24}$Mg are shown.  We have restricted
our  study to  the Ledoux  case because  the preclusion  of convection
implies a stronger heating of  the central region, due to the relative
inefficiency   of  the   radiative-conductive  transport   of  energy.
Evolutionary calculations  performed with the  Schwarzschild criterion
(Fig.~2)  confirm this  point,  even taking  into  account the  mixing
induced by the convective transport. As  it can be seen in Fig.~1, the
higher the abundance of  $^{24}$Mg, the higher the central temperature
achieved during the phase of electron capture on this isotope.

\begin{figure}
\vspace*{8.0cm}
\includegraphics{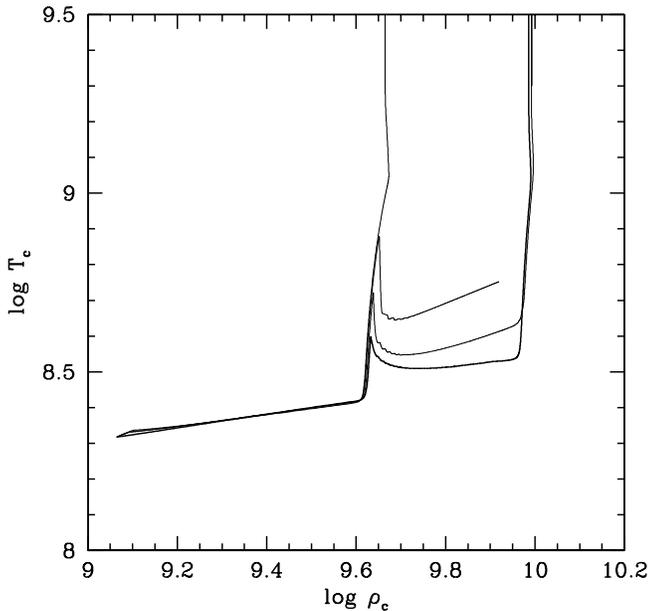}
\caption{Evolutionary sequences of  an ONe electron--degenerate object
         for  increasing abundances  of $^{24}$Mg.  The  abundances of
         $^{24}$Mg  are, respectively,  3\%,  10\%, 15\%  and 25\%  by
         mass.   The  model   with   $X(^{24}$Mg)=0.15  undergoes   an
         off-center  explosion and  the  model with  $X(^{24}$Mg)=0.25
         produces  a low  density explosion,  leading to  the complete
         disruption of the degenerate core.}
\end{figure}

\begin{table}[t]
\begin{center}
\begin{tabular}{ccc}
\hline
\hline
$X(^{24}$Mg) & $T_{\rm max,c}$ & $\rho_{\rm ign,c}$\\
 & (10$^8$ K) & (10$^9$ g cm$^{-3}$)\\
\hline
0.00 & $-$   & 9.410 \\
0.01 & 3.282 & 9.412 \\
0.02 & 3.561 & 9.422 \\
0.03 & 3.841 & 9.431 \\
0.04 & 3.957 & 9.682 \\
0.06 & 4.605 & 9.847 \\
0.08 & 5.261 & 10.221\\
0.10 & 5.957 & 10.409 \\
0.15 & 7.484 & 8.310$^\star$ \\
0.20 & 9.118 & 5.228$^\star$ \\
0.25 & 20.000 & 4.514 \\
\hline
\hline
\end{tabular}
\caption{Ignition  density  and  maximum central  temperature  reached
         during the  captures on $^{24}$Mg and  $^{24}$Na, for several
         initial  abundances of  $^{24}$Mg.  The  cases  whose central
         densities   are  marked  with   an  asterisk   correspond  to
         off-centre  explosions. In all  the calculations,  the Ledoux
         convective criterion has been adopted.}
\end{center}
\end{table}

Table 1 shows the net effect of increasing the abundance of $^{24}$Mg,
both for the  maximum central temperature and for  the central density
at ignition.  The  first trend obviously appearing in  Table 1 is that
failed explosions  lead to higher  collapse densities ---  see Fig.~1.
This is due  to the fact that electron  captures proceed in conditions
of constant $\rho Y_{\rm e}  \sim n_{\rm e}$, and hence higher amounts
of $^{24}$Mg  imply larger  reductions of $Y_{\rm  e}$.  Consequently,
the  collapse density  increases  from $9.41\times  10^9$~g~cm$^{-3}$,
when there  is no $^{24}$Mg in  the degenerate core,  to about $10.409
\times 10^9$~g~cm$^{-3}$,  when the amount of $^{24}$Mg  is 10\%.  The
minimum  abundance of  $^{24}$Mg  for which  NSE  develops during  the
electron capture  phase is  about 15\%, and  the explosion  is clearly
off-center.   The  reason is  simple:  during  the  phase of  electron
capture, part of  the energy generated is released  on the mass shells
above  that in which  the electron  capture reactions  are proceeding;
thus, the temperature of these shells is constantly increasing.  There
is also a second effect: a reduction in $Y_{\rm e}$ increases the rate
of contraction and, thus,  the gravitational enery release, leading to
a still higher  temperature of the material. This  can be readily seen
in Fig.~1, in  which the slope of the evolutionary  track in the $\log
(\rho_{\rm c})-\log (T_{\rm c})$ plane becomes steeper with increasing
abundances of $^{24}$Mg.  Eventually,  the energy generation in one of
these preheated shells  is able to drive the  temperature to the point
at  which  oxygen  burning   can  be  self-sustained,  and,  in  fact,
accelerated to produce the transition to NSE.  For even larger amounts
of $^{24}$Mg the  mass shell in which the  off-center explosion ensues
is located  closer to the  center.  The explanation is,  again, rather
straightforward,  as  the  preheating  for  increasing  abundances  of
$^{24}$Mg is more extreme and  the change in temperature caused by the
electron captures on $^{24}$Mg is also correspondingly higher.

\begin{figure}
\vspace*{8.0cm}
\includegraphics{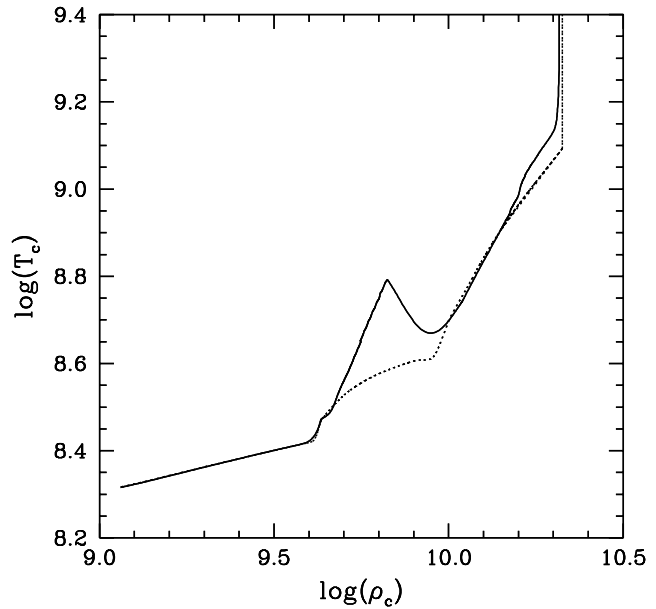}
\caption{Evolution of  an ONe object with  the Schwarzschild criterion
         (solid line). The dotted line  corresponds to a case in which
         convective mixing is inhibited.  In the case with mixing, the
         effects  of a  convective Urca shell are  evident.   The  peak
         corresponds to the  point where the nuclides of  mass 24 able
         to  capture electrons  (namely $^{24}$Mg  and  $^{24}$Na) are
         exhausted throughout the convective core.}
\end{figure}

A  central  explosion  occurs   only  when  the  amount  of  $^{24}$Mg
approaches  $\sim  25\%$, which  is  not  credible  given the  current
nuclear reaction rates relevant for the production of $^{24}$Mg during
the  carbon burning phase.   The same  can be  said about  the minimum
abundance  needed to  produce off-center  explosions: it  is extremely
unlikely ---  and many things  would need to change  significantly, in
particular the  rates of nuclear  reactions during carbon  burning ---
that $^{24}$Mg could by itself induce the heating to NSE in a electron
degenerate core at low density.

\section{Explosions induced by $^{12}$C}

\begin{figure}
\vspace*{8.0cm}
\includegraphics{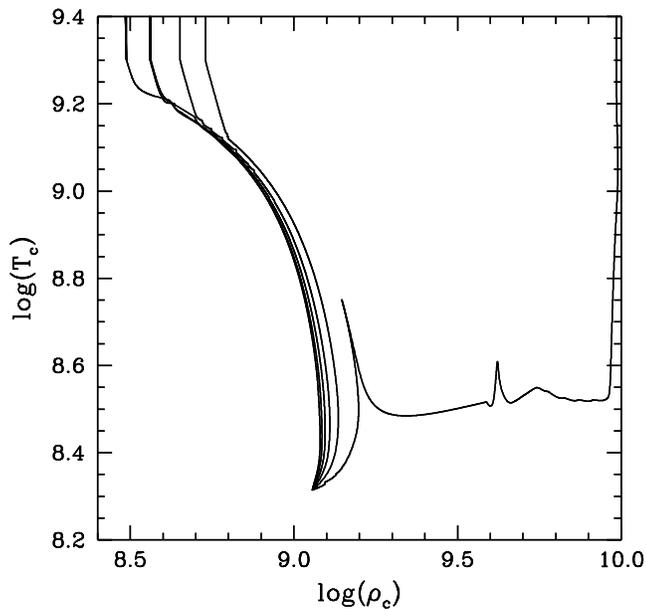}
\caption{Evolution on the  log($\rho_{\rm c}$)--log(T$_{\rm c}$) plane
         for the  models incorporating abundances  of $^{12}$C ranging
         from 0.01  to 0.06. The only  model that fails  to produce an
         explosion  at  moderate densities  is  that corresponding  to
         $X(^{12}$C)=0.01.   For all  these  calculations, the  Ledoux
         criterion has been adopted.}
\end{figure}

Recent  models (Dom\'{\i}nguez  et al.   1993; Ritossa  et  al.  1996;
Gil--Pons  \& Garc\'{\i}a--Berro 2001,  2002) have  convincingly shown
that carbon burning can be  incomplete in degenerate cores. The amount
of remaining carbon depends somewhat on the detail of the calculations
but  is   typically  of  the   order  of  $\la   1\%$.   Nevertheless,
Dom\'{\i}nguez et  al. (1993)  found a central  region of  $\sim 0.2\,
M_{\sun}$ with a  high carbon content, of the order  of 25\%, that may
eventually lead  to an early explosion,  and hence to  a disruption of
the  electron-degenerate core.   More specifically,  Dom\'{\i}nguez et
al.   (1993) followed  the evolution  of  a $10\,M_{\sun}$  star in  a
binary system.  The extreme  dependence of neutrino cooling on density
was  able to  cool down  the center  of the  star to  temperatures low
enough to  preclude the onset  of central carbon burning.  Although it
has been claimed that this result could be to due the roughness of the
numerical procedure  adopted in  Dom\'\i nguez et  al.  (1993)  --- in
particular to the poor spatial resolution  of their mesh --- or to the
very  high mass-loss  rate adopted,  this is  a point  that undoubtely
deserves more attention, since carbon burning in degenerate conditions
is prone to develop into  the explosive regime, even when this element
is  found  in  only  relatively  small amounts.   Woosley  (1986)  has
estimated  that, in the  interval $T_8$=2.5--7.5,  for each  0.12\% of
carbon burned the temperature increases by 10$^8$~K.

\begin{figure}
\vspace*{8.0cm}
\includegraphics{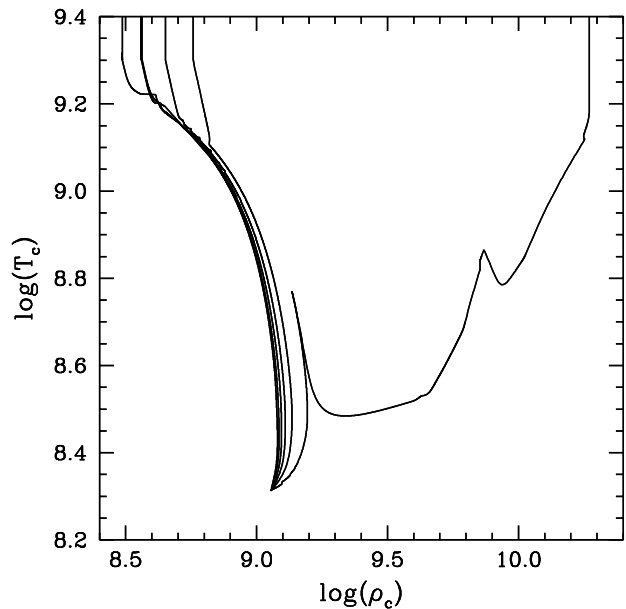}
\caption{Same  as figure  3,  but using  the Schwarzschild  convective
         criterion.  Only  the model  with  $X(^{12}$C)=0.01 fails  to
         trigger an explosion at moderate densities.}
\end{figure}

In this case,  given that there is a difference on  the atomic mass of
$^{12}$C  and its  daughter  nuclei (namely,  $^{16}$O, $^{20}$Ne  and
$^{24}$Mg), the  Ledoux criterion must  include an additional  term to
take it into account.  The expression employed throughout this work is
then

\begin{eqnarray}
\nabla_{\rm L} \equiv \nabla_{\rm ad} &-&
\left[ \left(\frac{\partial \ln P}{\partial \ln Y_{\rm e}}\right)_T\Bigg/
\left(\frac{\partial \ln P}{\partial \ln T}\right)_{Y_{\rm e}} \right]
\nabla_{Y_{\rm e}} \nonumber \\
&-& \left(\frac{\partial \ln Y_{\rm i} /\partial r}
{\partial \ln P / \partial r}\right)
\end{eqnarray}

Given the  strong dependence of  the carbon burning reaction  rates in
density and temperature,  it is expected that the  ignition will ensue
in  a very  limited region  of the  core. Hence,  even when  using the
Ledoux  criterion, convection will  develop at  the very  beginning of
carbon burning. This will induce strong mixing on the timescale of the
convective turnover.  Hence, fresh  material will replenish the carbon
content  of  the burning  region,  effectively  increasing the  energy
generation. Current  evolutionary calculations predict  a small amount
of unburnt  $^{12}$C distributed more  or less homogeneously  within a
substantial fraction of the core, with slightly higher abundances near
the center  and, consequently, the outcome of  these calculations will
depend sensitively on the adopted convective criterion.

\begin{table*}[t]
\begin{center}
\begin{tabular}{ccccc}
\hline
\hline
$X(^{12}$C) & Convection  &  $T_{\rm max}(^{12}$C)&
$\rho_{\rm ign}$(NeO) & Outcome\\
  &criterion & ($10^8$ K) & ($10^9$ g cm$^{-3}$) & \\
\hline
0.01 &  L  &  5.645 &  9.671 & C\\
     &  S  &  5.895 & 18.643 & C\\
     & $-$ &  4.242 &  9.663 & C\\
0.02 &  L   &  20.00 & 0.537  & E\\
     &  S   &  20.00 & 0.537  & E\\
     &  $-$ &  6.901 & 9.653  & C\\
0.03 &  L   &  20.00 & 0.450  & E\\
     &  S   &  20.00 & 0.448  & E\\
     &  $-$ &  9.084 & 9.686  & C\\
0.04 &  L   &  20.00 & 0.365  & E \\
     &  S   &  20.00 & 0.363  & E \\
     &  $-$ &  11.22 & 1.200$^\star$ & E \\
0.05 & L   &  20.00 & 0.366  & E \\
     & S   &  20.00 & 0.365  & E \\
     & $-$ &  20.00 & 1.165  & E \\
0.06 & L   &  20.00 & 0.308  & E \\
     & S   &  20.00 & 0.307  & E \\
     & $-$ &  20.00 & 1.108  & E \\
\hline
\hline
\end{tabular}
\caption{Models with variable $^{12}$C abundance.  In the table, ``L''
         stands for  the Ledoux criterion, while ``S''  stands for the
         Schwarzschild  criterion; ``C''  denotes  collapse and  ``E''
         disruption.   For carbon abundances  over 2\%,  all realistic
         models imply  the complete disruption  of the star.  The case
         with  a 4\%  abundance  of carbon,  and  with the  convection
         artificially inhibited (marked by an asterisk) corresponds to
         an off-center explosion.}
\end{center}
\end{table*}

In order to determine the minimum amount of remaining carbon needed to
completely disrupt  the core at  low densities, we  have parameterized
its abundance from  1\% to 6\%, a broad enough  range. We have adopted
both the Ledoux and  the Schwarzschild criteria.  Also, to investigate
the worst possible case, we have artificially inhibited the convective
motion and energy transport.  In  this case carbon burning proceeds in
a strictly local  mode, making the onset of  NSE more difficult.  This
could be similar  to what would happen if  the thermonuclear timescale
were  much  shorter  than  the convective  turnover  and  evolutionary
timescales.

The first set of calculations is shown in Fig.~3, for which the Ledoux
criterion was  used. As it  can be seen,  the only model for  which an
early   explosion   is  avoided   is   that   with  carbon   abundance
$X(^{12}$C$)\sim 0.01$. In Fig.~4 the evolution in the $\log(\rho_{\rm
c})-\log(T_{\rm c})$ plane is shown for the set of models in which the
Swarzschild  criterion  was  adopted.   Again, only  the  models  with
$X(^{12}$C$)\la 0.01$ do not lead  to an early explosion.  Finally, in
Fig.~5 the results obtained when convection was artificially inhibited
are shown. In this case, those models in which the carbon abundance is
$\ga 2\%$  give an explosion at moderate  densities. For completeness,
our results are  summarized in Table 2, where  the maximum temperature
attained during  carbon burning (fourth column),  the ignition density
(fith  column) and  the outcome  of the  evolution (sixth  column) are
shown  as  a function  of  the  carbon  abundance and  the  convective
criterion. Table 2  shows that in the cases  where convection is taken
into account, the  minimum abundance of $^{12}$C should  be as high as
1.5\%, only a factor of 2 higher than the abundance presently found in
most   calculations.    Even  when   the   convective  processes   are
artificially inhibited, the amount of $^{12}$C required to achieve NSE
at  low  densities increases  to  4\%  ---  and the  explosion  occurs
slightly off-center.  It  is clear that the presence  of small amounts
of  unburned $^{12}$C  could be  a possible  channel for  the complete
disruption  of  ONe degenerate  objects  at  very  low density  unless
non-physically  consistent situations are  artificially forced.  It is
also not  surprising that a  model adopting the  $^{12}$C distribution
obtained by Dom\'{\i}nguez  et al.  (1993) results in  an explosion at
low   density.   Nevertheless,  Rayleigh-Taylor   instabilities  would
rapidly mix  the $^{12}$C-rich  matter with the  rest of the  core. In
this  case, the  abundance of  $^{12}$C  is already  within the  range
studied in the present work (about a 4.5\%).

Given the low density at which NSE proceeds, it is likely that some of
the reactions  will be rapidly quenched, thus  producing an incomplete
burning and a corresponding  lower fraction of Fe-peak elements. While
this is not compatible with  classical type Ia SNe nucleosynthesis, it
could be an alternative explanation  for subluminous type Ia SNe, like
SN1991bg (Ruiz-Lapuente et al.   1993). Clearly, this outcome deserves
further consideration, which is beyond the scope of this paper.

\section{Effects of the Urca pairs $^{23}$Na-$^{23}$Ne and $^{25}$Mg-$^{25}$Na}

Recent evolutionary  calculations including detailed  nuclear reaction
networks have found that noticeable amounts of $^{23}$Na and $^{25}$Mg
appear after the carbon burning  phase --- see, for example, Gil--Pons
\&    Garc\'{\i}a--Berro   (2001).     These    nuclides   have    low
electron-capture  energy thresholds ($Q=$4.376~MeV  and $Q=$3.833~MeV,
respectively) and  so they will start captures  even before $^{24}$Mg,
which  has a  higher energy  threshold. The  Urca processes  for these
nuclides     are     $^{23}$Na(e,$\nu)^{23}$Na(,e$\nu)^{23}$Ne     and
$^{25}$Mg(e,$\nu)^{25}$Na(,e$\nu)^{25}$Mg,  occurring at  densities of
$1.68\times   10^9$~g~cm$^{-3}$    for   $^{23}$Na   and   $1.17\times
10^9$~g~cm$^{-3}$ for  $^{25}$Mg. The rates for these  Urca pairs have
also been  taken from  Oda et al.   (1994).  Note that  throughout the
calculations  presented  in  this  section  Coulomb  corrections  were
applied  to   the  rates,  as  was  done   previously.   The  chemical
composition of the ONe core was $X(^{16}$O)=0.625, $X(^{20}$Ne)=0.262,
$X(^{23}$Na)=0.052,  $X(^{24}$Mg)=0.045,   and  $X(^{25}$Mg)=0.015  as
derived from Ritossa et al.  (1996).  Also note that for simplicity we
have excluded the small abundance of unburnt $^{12}$C and renormalized
the resulting abundances.

\begin{figure}
\vspace*{8.0cm}
\includegraphics{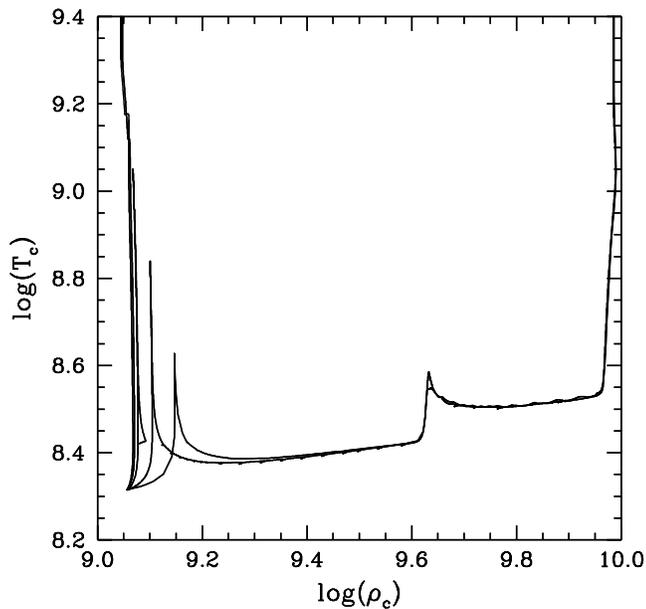}
\caption{Same  as figures 3  and 4,  but with  convection artificially
         inhibited.  In  this case, even  a 2\% abundance  of $^{12}$C
         fails to trigger a moderate density explosion.}
\end{figure}

As  it can  be seen  in Fig.~6,  the main  effect of  the  Urca pairs
consist of  small decreases  in $T_{\rm c}$  and in $Y_{\rm  e}$. The
later effect is  the only lasting one.  These Urca  pairs are never in
equilibrium, as the core  is slowly contracting and, hence, increasing
the  Fermi energy  of the  electrons.  Some  $10^9$ seconds  after 
formation of  the Urca  pairs, the rates  for electron-capture  on the
involved  nuclides are  much  higher than  the $\beta$--decay 
rates.   Eventually,  all the  $^{23}$Na  becomes  $^{23}$Ne, and  the
$^{25}$Mg becomes $^{25}$Na, with a  total reduction on $Y_{\rm e}$ of
0.00286, which  translates into a slightly higher ignition  density for
NeO burning.

\begin{figure}
\vspace*{8.0cm}
\includegraphics{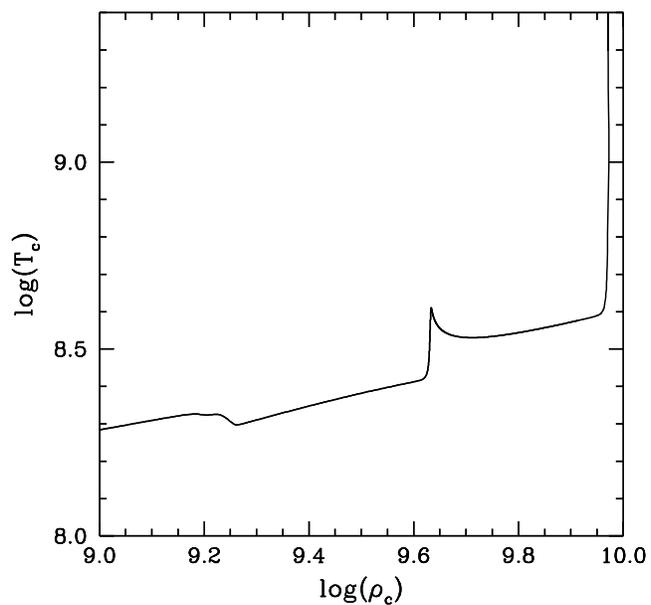}
\caption{Evolution in  the $\log(\rho_{\rm c})-\log(T_{\rm  c}$) plane
         of  an ONe  electron--degenerate object  with  the abundances
         given  by Ritossa  et al.   (1994).  The  effect of  the Urca
         pairs  involving $^{23}$Na  and  $^{25}$Mg is  apparent as  a
         sudden decrease in $T_{\rm c}$ near $\log\rho_{\rm c}$=9.2.
         The convective criterion adopted was that of Ledoux.}
\end{figure}

When using the Schwarzschild  criterion, convection appears during the
electron  captures   on  the  nuclides  of  mass   24  ($^{24}$Mg  and
$^{24}$Na).  Due to the limited extent of the convective region, there
is no convective Urca for $^{23}$Ne nor for $^{25}$Na, since the upper
layer  of  the  convective  region  is  well  above  the  density  for
significant  rates of $\beta$  decay from  these nuclides.   Thus, the
only durable effect of $^{23}$Na and $^{25}$Mg on the evolution of our
objects is a  small reduction in $Y_{\rm e}$  (from the original value
of 0.4981 to 0.4952).  Hence,  due to this slight reduction in $Y_{\rm
e}$,  the ignition  density for  NeO burning  is $9.493\times  10^9$ g
cm$^{-3}$,  clearly above the  collapse threshold  given by  Timmes \&
Woosley (1992, 1994).

\section{Conclusions}

The  final fate  of  pure  ONe electron  degenerate  objects seems  at
present well established:  they will collapse to a  neutron star after
electron  captures  on  $^{20}$Ne  at  a  density  of  about  $9\times
10^9$~g~cm$^{-3}$. The  main goal of  this work has been  to elucidate
whether the minor  species present in ONe cores  --- namely, $^{12}$C,
$^{23}$Na, $^{24}$Mg and  $^{25}$Mg --- can open a  channel to produce
an  explosion at moderate  densities (about  $10^9$~g~cm$^{-3}$) which
would lead to the complete disruption of the star.

We have found  that the amount of $^{24}$Mg required  to produce a low
density explosion is about one order of magnitude above the results of
current,  state-of-the-art evolutionary  calculations.   Hence, it  is
very unlikely that electron  captures proceeding on $^{24}$Mg (and its
daughter nucleus $^{24}$Na)  can drive the temperature of  the star to
the   point   of   igniting    the   NeO   burning   at   $\rho   \sim
4.5\times10^9$~g~cm$^{-3}$.  Even  in the (improbable)  case that this
actually happened, this density is near the current accepted threshold
for    gravitational    collapse     (Timmes    \&    Woosley    1992,
1994). Consequently, if there is  not a very significant change in the
reaction rates  leading to the  production of $^{24}$Mg,  this nuclide
can  be  discarded  as  a  possible  trigger  of  a  moderate  density
explosion.

On the contrary, even small amounts of unburned $^{12}$C are enough to
drive a  moderate to  low density explosion.   The reason  is twofold.
First, the  energy yield  is much higher  for carbon burning  than for
electron captures on the $A=24$  nuclei.  Secondly, there is much more
$^{12}$C available,  since the ignition  is very local and  can induce
convective  motion that leads  to mixing  that carries  fresh $^{12}$C
into the burning region.  In this way, even seemingly small abundances
of $^{12}$C (about a 1.5\%)  can produce an explosion at $\rho_{\rm c}
\sim  1\times10^9$~g~cm$^{-3}$,  which  would  lead  to  the  complete
disruption of the star.  This potential channel to complete disruption
would give rise to  a thermonuclear supernova.  The characteristics of
such a  kind of thermonuclear supernova  are beyond the  scope of this
work, but  could explain in part  why there seems to  be a correlation
between  spiral arms and  Type Ia  supernovae (Bartsunov,  Tsvetkov \&
Filimonova 1994; McMillan \&  Ciardullo 1996). Finally, the effects of
the Urca pairs generated by  $^{23}$Na and $^{25}$Mg are minor.  Apart
from a small reduction in  the temperature, the only lasting effect is
a small  decrease in the electron  mole number $Y_{\rm  e}$. This will
result in a  slightly higher density for the  electron captures on the
nuclei of $A=24$ and $A=20$.

We  note that  multidimensional simulations  of convection  indicate a
tendency to  increase the  amount of convective  mixing and  to induce
rotational  mixing  to  an  extent  that  could  slightly  modify  our
conclusions.  Many  efforts are  currently being invested  in modeling
carbon   burning   under    strongly   degenerate   conditions   using
multidimensional codes  --- see,  for instance, Reinecke,  Niemeyer \&
Hillebrandt (2003) and R\"opke,  Hillebrandt \& Niemeyer (2004a,b) for
recent  developments  in  this  field.   A  time--dependent  numerical
procedure  for mixing  (necessary  for coupling  core contraction  and
convection) could slightly change  the details of our conclusions, but
it is unlikely to cause a qualitative change --- that is, transforming
a  collapse  into  an  explosion  or vice  versa.   Until  a  complete
treatment of multidimensional convection and burning is available, our
results can be regarded as  a fair approximation to the real evolution
of this type of star.

\begin{acknowledgements}
Part of this work was supported by the MCYT grant AYA04094--C03-01, by
the European Union  FEDER funds, and by the CIRIT.  We would also like
to  acknowledge the  invaluable advice  of J. Isern and  J.  Labay who
largely contributed  through his support, suggestions  and comments to
improve the manuscript.
\end{acknowledgements}

\end{document}